\documentclass{ifacconf}

\usepackage{natbib}        %
\usepackage{amsmath,color} 
\usepackage{amssymb} 
\usepackage{amsxtra}
\usepackage{arydshln}
\usepackage{amsfonts} 
\usepackage[mathscr]{eucal}
\usepackage{mathrsfs} 
\usepackage{setspace} 
\usepackage{url}
\usepackage{graphicx} 
\usepackage{multirow}
\usepackage{palatino,epsfig,fleqn}
\usepackage{accents}

\newcommand{\hs}{\hspace{4mm}}

\newcommand{\IR}{\mathbb{R}}
\newcommand{\IH}{\mathbb{H}}
\newcommand{\IL}{\mathbb{L}}
\newcommand{\IS}{\mathbb{S}}

\definecolor{orangered}{rgb}{0.8,0.1,0.1}
\definecolor{grassgreen}{rgb}{0.1,0.65,0.25}

\newcommand{\SH}{\color{black}}
\newcommand{\yh}{\color{black}}
\newcommand{\ck}{\color{black}}

\DeclareMathAlphabet{\matheur}{U}{eur}{m}{n}
\DeclareMathAlphabet{\matheurb}{U}{eur}{b}{n}
\DeclareMathAlphabet{\matheus}{U}{eus}{m}{n}
\DeclareMathAlphabet{\matheuf}{U}{euf}{m}{n}

\newcommand{\RHinf}{\mathbb{RH}_\infty}
\newcommand{\Hinf}{\mathbb{H}_\infty}
\newcommand{\RLinf}{\IR \IL_\infty}

\newcommand{\AP}{{\cal AP}}
\newcommand{\F}{\mathcal{RF}}

\newtheorem{lemma}{\bfseries Lemma}
\newtheorem{theorem}{\bfseries Theorem}

\def\QED{~\rule[-1pt]{5pt}{5pt}\par}
\newenvironment{proof}{{\em Proof.}}{{ \hfill \QED}}
\newtheorem{remark}{\bfseries Remark}

\usepackage{fancyhdr}
\fancypagestyle{firstpage}{
\fancyhead{}
\fancyfoot{}
\lfoot{\scriptsize 
\copyright 2023 Chung-Yao Kao, Shinji Hara, Yutaka Hori, Tetsuya Iwasaki, Sei Zhen Khong. This work has been accepted to IFAC for publication under a Creative Commons Licence CC-BY-NC-ND. }

}

\begin{document}
\begin{frontmatter}

\title{On Phase Change Rate Maximization with Practical Applications~\thanksref{footnoteinfo}} 

\thanks[footnoteinfo]{This work was supported in part by the National
Science and Technology Council of Taiwan, under grant MOST 110-2221-E-110-047-MY3. }

\author[First]{C.-Y. Kao} 
\author[Second]{S. Hara} 
\author[Third]{Y. Hori}
\author[Fourth]{T. Iwasaki}
\author[Fifth]{S. Z. Khong}

\address[First]{Dept. of Electrical Engineering, 
National Sun Yat-Sen University, Taiwan. (e-mail: cykao@mail.ee.nsysu.edu.tw)}
\address[Second]{Global Scientific Information and Computing Center, Tokyo Institute of Technology, Japan. (e-mail: shinji\_hara@ipc.i.u-tokyo.ac.jp)}
\address[Third]{Applied Physics and Physico-Informatics, Keio University, Japan. (e-mail: yhori@appi.keio.ac.jp)}
\address[Fourth]{Dept. of Mechanical and Aerospace Engineering, University of California at Los Angeles, USA. (e-mail: tiwasaki@ucla.edu)}
\address[Fifth]{Independent Researcher. (email: szkhongwork@gmail.com)}

\begin{abstract}                %
  We recapitulate the notion of phase change rate maximization and demonstrate the usefulness of its solution on analyzing the 
  robust instability of a cyclic network of multi-agent systems subject to a homogenous multiplicative perturbation. Subsequently, 
  we apply the phase change rate maximization result to two practical applications. The first is a magnetic levitation system, while 
  the second is a repressilator with time-delay in synthetic biology. 
\end{abstract}


\end{frontmatter}

{
\thispagestyle{firstpage}
\section{Introduction}
\label{sec:Intro}

Robustness against model uncertainties for feedback systems has been recognized as one of the important issues 
in control theory from the practical application viewpoint over forty years since the 1980s. 
The most typical and successful theory is the $H_\infty$ control which includes robust stability and robust stabilization 
against norm-bounded dynamic uncertainties. See e.g.,~\citep{zhou:96book} and the references therein. 

A counterpart of the robust stability analysis is the so-called {\it ``robust instability analysis''} for nominally 
unstable feedback systems, and the problem is to find a stable perturbation with the smallest $H_\infty$-norm 
which stabilizes the system. A practical motivation of the analysis is maintaining nonlinear oscillations caused by instability of an equilibrium
point for dynamical systems arising in neuroscience and synthetic biology.  See~\citep{HIH:CDC2020} and~\citep{HIH:Automatica2021} 
for applications to the FitzHugh-Nagumo neuron model and repressilator model, respectively.  

The instability analysis problem is closely related to the strong stabilization, i.e., stabilization by a stable controller
\citep{Youla:Automatica1974, Zeren:Automatica2000, Ohta:TSICE2001}. Actually, it is equivalent to strong stabilization by a minimum-norm controller. The problem is extremely difficult due to the following two reasons: (i) non-convexity nature of minimum-norm controller
synthesis and (ii) no upper bound on the order of stable stabilizing controllers. In other words, the robust instability analysis is 
similar to the robust stability analysis in terms of the problem formulation, but it is quite different technically and much more
challenging as optimization problems.

Recently, the authors proposed a new optimization problem, which we call the {\it ``Phase Change Rate Maximization Problem''} in order 
to provide an almost complete solution to the small-gain-type condition for the robust instability analysis for some classes of systems with one or two unstable 
poles~\citep{HKSZIH:arXiv2022}. The problem is to find a {\ck stable} real-rational transfer function such that its peak gain occurs 
at a given frequency $\omega_p$ with a prescribed phase value, and the phase change rate (PCR) at $\omega_p$ is the maximum among 
those satisfying the constraints. The essential idea behind is the following. One of the key factors for the difficulty of robust 
instability analysis is that we cannot detect the transition from instability to stability by the presence of a pole on the imaginary axis 
(which successfully characterizes the transition in the opposite direction, making the robust stability analysis tractable).
Hence we need an additional criterion for the transition. It turned out, roughly speaking, that the positivity of the PCR of the loop 
transfer function at the peak gain frequency is an indication of the instability-to-stability transition for certain systems.
The aforementioned paper showed that the maximum PCR is attained by a first-order all-pass function and derived conditions under 
which the exact robust instability analysis is possible in terms of the PCR.  

The purpose of this paper is twofold. The first purpose is to supplement the theoretical results in~\citep{HKSZIH:arXiv2022} by a more
comprehensive example than those in the reference and illustrate how the PCR plays an important role for the exact robust 
instability analysis. The class of systems is given as cyclic networks of homogeneous agents,
where by changing the number of agents we can treat a variety of situations with respect to the location of 
stable and unstable complex poles with relatively small dampings. We focus especially on the relationship between the sign of 
the PCR and the stable/unstable poles which are fairly close to the imaginary axis and represent under what situation we can get 
the exact result. The second purpose is to show that the PCR condition derived in~\citep{HKSZIH:arXiv2022} works well for two practical applications, namely (i) a minimum-norm strong stabilization for magnetic levitation systems and 
(ii) an exact robust instability analysis for the repressilator with time delay. 
The target systems of the former and the latter cases are in 
$\mathcal{G}_1^0$ (one unstable pole with the peak gain attained at zero frequency) and 
$\mathcal{G}_2^\#$ (two unstable poles with the peak gain attained at non-zero frequency) , 
respectively, for which we can get the exact results. 
This means that the theoretical foundation in~\citep{HKSZIH:arXiv2022} can be practically useful 
although the class of applicable systems may appear restricted.  

The remainder of this paper is organized as follows. Section~\ref{sec:PCRmaximization} is devoted to a brief summary of the PCR
maximization problem presented in~\citep{HKSZIH:arXiv2022} and an illustrative example.  Section~\ref{sec:PracticalApp} provides two practical applications. Section~\ref{sec:Concl} summarizes the contributions of this paper and addresses some future research directions.
}

{\bf Notation and Terminology:}
The set of real numbers is denoted by $\IR$. 
$\Re(s)$ and $\Im(s)$ denote the real and imaginary parts of a complex number $s$, respectively. 
The set of proper real rational functions of one complex variable $s$ is denoted by $\IR_p$.
Let $\IL_{\infty}$ denote the set of functions that are bounded on the imaginary axis $j\IR$. 
The subset of $\IL_{\infty}$ which consists of real rational functions bounded on $j\IR$ is 
denoted by $\RLinf$. The stable subsets of $\IL_{\infty}$ and $\RLinf$ are denoted by 
$\Hinf$ and $\RHinf$, respectively. 
The norms in $\IL_{\infty}$ and $\Hinf$ are denoted by
$\| \cdot \|_{L_\infty}$ and $\| \cdot \|_{H_\infty}$, respectively.
The open (closed) left and right half complex planes are abbreviated as OLHP (CLHP) and ORHP (CRHP), respectively.

The following terminology will be used for a rational function ${h \in \IR_p}$ throughout the paper:
$h$ is called ``stable'' (or ``exponentially stable'') if all the poles of $h$ are in the OLHP;   
``marginally stable'' if all the poles of $h$ are in the CLHP and any pole of $h$ on the imaginary axis is simple; 
``unstable'' (or ``exponentially unstable'') if at least one of the poles of $h$ is in the ORHP.  

\section{Phase Change Rate Maximization}
\label{sec:PCRmaximization}

In this section, we introduce the PCR maximization problem, and motivate the problem 
by instability analysis and strong stabilization. 
\subsection{Problem Formulation}
Given $\omega_p>0$ and $\theta_p\in[0,2\pi)$, we consider the following ``phase change rate'' maximization problem
\begin{align}
\label{eq:prob_org}
\sup_{f\in\RHinf}~\theta_f'(\omega_p) ~~\mbox{s.t.} ~~
\|f\|_{H_\infty}=|f(j\omega_p)|, ~~ \theta_f(\omega_p)=\theta_p,
\end{align}
where $\theta_f(\omega)$ denotes the phase angle of $f(j\omega)$, and $\theta_f'(\omega)$ is its derivative. In other words, 
we seek a function $f$ from $\RHinf$, whose $\Hinf$-norm occurs at frequency $\omega_p$ and phase 
at $\omega_p$ is constrained to be $\theta_p$, and has the maximal ``phase change rate'' among all
functions which satisfy the same constraints. Such problem arises from robust instability analysis and minimum-norm strong 
stabilization as explained below. 

Consider a positive feedback system with {\ck a} loop-transfer function
$g(s)\delta(s)$; i.e., the characteristic equation of the system is given by
$1 - g(s)\delta(s) = 0$, where $g(s)$ denotes the nominal {\ck part} which
belongs to a class of unstable systems defined by
\begin{equation*} %
\hspace{-0.4cm}
 {\cal G} := \{  g \in \IR\IL_\infty \; | \;  g \; \mbox{is strictly proper and unstable} \}  
\end{equation*}
and $\delta(s)$ represents a real-rational dynamic perturbation.
The robust instability radius (RIR) for $g \in {\cal G}$ {\ck with respect to $\delta\in \RHinf$}, 
denoted by $\rho_*(g) \in \IR$,
is defined as the smallest magnitude of the perturbation that internally stabilizes the system:
\begin{equation} \label{rho}
\rho_*(g) := \inf_{\delta\in\IS(g)}~\|\delta\|_{H_\infty},
\end{equation}
where $\IS(g)$ is the set of real-rational, proper, stable transfer functions
internally stabilizing $g$; {\it i.e.},
\begin{equation*} %
\begin{array}{r}
\hspace{-1mm}
\IS(g) := \{ \delta\in\RHinf:~ \delta(s)g(s)=1 ~ \Rightarrow ~ \Re(s)<0, ~~\\
\delta(s)=0, ~ \Re(s)>0 ~ \Rightarrow ~ |g(s)| < \infty ~ \}.
\end{array}
\end{equation*}

The optimization problem stated in~\eqref{rho} is identical to the so-called ``minimum-norm strong stabilization''
problem for a given (unstable) plant $g$, where the minimum-norm controller sought is required to be stable itself. 
It is noticed from the well known result on strong stabilizability in \citep{Youla:Automatica1974} 
that $\rho_*(g)$ is finite if and only if the Parity Interlacing Property 
(PIP) %
is satisfied, 
{\it i.e.,} the number of unstable real poles of $g$ between any pair of real zeros 
in the closed right half complex plane (including zero at $\infty$) is even. 
Consequently, the class of systems of our interest is defined as
\begin{align*}
\begin{split}
\mathcal{G}_n := \{g\in{\cal G} \; | \;  & g \; 
 \mbox{has $n$ unstable poles and} \\
&  \mbox{satisfies the PIP condition} \}, 
\end{split}%
\end{align*}
where $n$ is a natural number. Let $g \in {\cal G}$ be given. We have the following lower bound 
for $\rho_*(g)$~(see \citep{HIH:Automatica2021})   
\begin{equation*} %
 \rho_* (g) \geq ~ 1/\|g\|_{L_\infty} , \hs \|g\|_{L_\infty} := \sup_{\omega \in \IR} |g(j\omega)| . 
\end{equation*} 
When $\rho_*(g)$ is exactly equal to its lower bound $1/\|g\|_{L_\infty}$, we say 
$g$ has \emph{the exact RIR}. It has been shown in~\citep{HIH:Automatica2021} that, if $f$ 
with $\|f\|_{H_\infty}=1/\|g\|_{L_\infty}$ marginally stabilizes $g$ with a single pair
of poles on the imaginary axis, then $g$ has the exact RIR. 
Moreover, based on an extended version of the Nyquist criteria, necessary and sufficient conditions 
were derived in~\citep{HKSZIH:arXiv2022} for marginal stabilization of $g$, which in turn are 
sufficient conditions for obtaining the exact RIR of $g$. As a part of the necessary and sufficient 
condition for $f$ with $\|f\|_{H_\infty}=1/\|g\|_{L_\infty}$ to marginally stabilize $g$, the open-loop
transfer function $gf$ must satisfy the following loop-gain and PCR conditions:
\begin{align*}
g(j\omega_p)f(j\omega_p) = 1, \ \ \theta_{gf}'(\omega_p) = \theta_{g}'(\omega_p) + \theta_{f}'(\omega_p)>0,
\end{align*}
where $\omega_p$ is the 
frequency where the $L_\infty$-gain of $g$ occurs. Searching for such an $f$ boils down to solving a
PCR optimization problem of the form described in~\eqref{eq:prob_org}, where 
the phase $\theta_f(\omega_p)$ is constrained to $-\theta_g(\omega_p)$ (and the magnitude of $f$
at $\omega_p$ is irrelevant to PCR optimization, as positive scaling of $f$ will not 
change its phase or phase change rate). The solution of the problem provides a
tight condition for $g$ to be marginally stabilizable. 
In the next subsection, we summarize the theoretical foundation in~\citep{HKSZIH:arXiv2022}.

\subsection{The Solution and its Application to Instability Analysis}

The PCR optimization in (\ref{eq:prob_org}) can be solved by first narrowing down the 
feasible set using the following sets of functions:
\begin{align*}
\F_{\omega_p, \theta_p}  := \{ &f  \in \IR\IH_\infty  : 1=\|f\|_{H_\infty}=|f(\omega_p)|, \\ 
& \hspace{3.7cm} \theta_f(\omega_p) = \theta_p \}. \notag 
\end{align*}
\begin{align*}
\mathcal{O}_{\omega_p, \theta_p} := \{ &f  \in \IR\IH_{\infty} : f \text{ is minimum phase, }\\
 &|f(j\omega_p)| = \|f\|_{H_{\infty}}, \text{ and } \theta_f(\omega_p) = \theta_p\}. \notag 
\end{align*}
\begin{align*}
\AP_{\omega_p,\theta_p}:= \{ &f  \in \IR\IH_{\infty} : |f(j\omega )|  = 1, \forall \omega , \\
 &|f(j\omega_p)| = \|f\|_{H_{\infty}}, \text{ and } \theta_f(\omega_p) = \theta_p\}. \notag 
\end{align*}
Note that the constraint on the magnitude of the $\Hinf$-norm of functions in $\F_{\bullet,\bullet}$ and 
$\AP_{\bullet,\bullet}$ bears no significance as explained previously. The constraint is placed for convenience only. 
The first result gives an upper bound on the PCR for functions in $\mathcal{O}_{\omega_p,\theta_p}$.

\begin{prop}
\label{thm:bound} 
Let $\theta_p \in (-\pi, \pi]$ and $f\in\mathcal{O}_{\omega_p, \theta_p}$ be given. If $\omega_p \not = 0$, 
then $\theta_f^\prime(\omega_p) \leq - \left|\theta_p/\omega_p\right|$. Moreover, if $\omega_p=0$, then $\theta_f'(\omega_p) \le 0$.
\end{prop}
Proposition~\ref{thm:bound} establishes that, for a stable minimum-phase system, its PCR at the 
peak-frequency (i.e., where the $\Hinf$-norm occurs) is always non-positive. Since any $\RHinf$ function can be
factorized as multiplication of an all-pass function and a minimum-phase function, Proposition~\ref{thm:bound}
suggests that the PCR maximization problem over the set $\F_{\bullet,\bullet}$ boils down to the problem
over the set $\AP_{\bullet,\bullet}$. This is indeed the case, as the following proposition states. 
\begin{prop}
\label{thm:PCR_Sol}
Given $\omega_p \not= 0$ and $\theta_{p}\in(-\pi,\pi]$ (mod $2\pi$), we have 
\begin{align*} %
\hspace{-3mm}
\sup_{f\in \F_{\omega_p,\theta_{p}}} \theta_f'(\omega_p) =  \sup_{f\in \AP_{\omega_p,\theta_{p}}} \theta_f'(\omega_p) = 
-\left|\sin(\theta_{p})/\omega_p\right|.
\end{align*}
Moreover, when $\theta_{p}\not\in\{0,\pi\}$, the supremum is attained by the first-order all-pass 
function of the form $f(s)=\frac{a-s}{a+s}$ or $f(s)=\frac{s-a}{a+s}$. 
When $\theta_{p}\in\{0,\pi\}$, the supremum is attained by a zeroth-order all-pass functions; 
i.e., $f(s)= 1$ or $f(s)= -1$. 
For $\omega_p=0$, the only feasible phase angles are $\theta_p\in\{0,\pi\}$ (mod $2\pi$). In 
this case,  
$$
\sup_{f\in \F_{0,\theta_{p}}} \theta_f'(0) =  \sup_{f\in \AP_{0,\theta_{p}}} \theta_f'(0) = 0.
$$
The supremum is attained by $f(s)= 1$ or $f(s)= -1$. 
\end{prop}

Using the solutions stated in Proposition~\ref{thm:PCR_Sol}, the following results were derived for 
two subclasses of $\mathcal{G}_n$ defined by
\begin{align*}
&\hspace{-0.5cm}\mathcal{G}_n^0:=\{g\in\mathcal{G}_n~|~  
\|g\|_{L_\infty}=|g(0)|>|g(j\omega)|~\forall\omega\neq 0\}, \\
&\hspace{-0.5cm}\mathcal{G}_n^{\#}:=\{g\in\mathcal{G}_n~|~  
\exists~\omega_p>0~\mbox{such that}~\notag \\
&\hspace{1cm}\|g\|_{L_\infty}=|g(j\omega_p)|>|g(j\omega)|~\forall\omega\neq\pm\omega_p
\}
\end{align*}
based on an extended Nyquist criterion~\citep{HKSZIH:arXiv2022}.
\begin{theorem} \label{thm:marginalstab} 
\hs
\begin{itemize}
\item[(I)]  
Given $g \in {\cal G}_n^0$,  
$g$ can be marginally stabilized by a stable system $f$ 
with $\|f\|_{H_\infty} = 1/\|g\|_{L_\infty} = 1/|g(0)|$ if and only if $n=1$ and $\theta_g'(0)>0$.
\item[(II)] Given $g \in {\cal G}_n^\#$ for which 
the peak gain occurs at $\omega_p$,
 $g$ can be marginally stabilized by a stable system $f$ 
with $\|f\|_{H_\infty} = 1/\|g\|_{L_\infty} = 1/|g(j\omega_p)|$ if and only if $n=2$ and
$\theta_g'(\omega_p)    >  \left|\sin(\theta_g(\omega_p))/\omega_p\right|$.
\end{itemize}
\end{theorem}
Note that the marginally stabilizing controllers for cases (I) and (II) can be 
taken as the zeroth-order and the first-order all-pass functions, respectively, as
suggested by Proposition~\ref{thm:PCR_Sol}. 

As marginal stabilization of a system guarantees the exact RIR for the system, 
Theorem~\ref{thm:marginalstab} immediately leads to sufficient conditions 
for attaining the exact RIR of systems in $\mathcal{G}_1$ and $\mathcal{G}_2$. 
Furthermore, necessary conditions can also be derived based on the following result, 
which gives a PCR condition on the loop-transfer function at the peak 
frequency when the closed-loop system has all its pole in the closed left half plane.
\begin{lemma}\cite[Lemma 5]{HKSZIH:arXiv2022}
\label{lemma:PS}
Given $\omega_c\ge 0$, an integer $n\ge 1$, and a transfer function $L\in\mathcal{G}_n$,  
consider the positive feedback system with loop transfer function $L$ satisfying the following condition
\begin{align*}
\begin{split}
&1=|L(j\omega_p)|=\|L\|_{L_\infty}, \\ 
&|L(j\omega)|<|L(j\omega_p)|, \forall \omega\not=\pm\omega_p.
\end{split}
\end{align*}
If the feedback system has all its poles in the CLHP, then $\theta_L'(\omega_p)\ge 0$. 
\end{lemma}
Based on Theorem~\ref{thm:marginalstab} and Lemma~\ref{lemma:PS},
we have necessary conditions and sufficient conditions for the exact RIR 
as follows. 

\begin{theorem}\label{thm:exactRIR}
Let $g\in{\cal G}$ be given. Suppose $g(j\omega)$ takes the peak gain at
$\omega_p$ and consider the exact RIR condition
\begin{equation} \label{exactRIR}
\rho_*(g) = 1/\|g\|_{L_\infty} = 1/|g(j\omega_p)|.
\end{equation}
\begin{itemize}
\item[(I)]  
Suppose $g \in {\cal G}_1^0$ and $\omega_p=0$. Then
\begin{align*}
\theta_g'(\omega_p)>0 
~~\Rightarrow~~ 
(\ref{exactRIR})
~~\Rightarrow~~ 
\theta_g'(\omega_p)\ge 0.
\end{align*}
\vspace{-0.15cm}
\item[(II)] Suppose $g \in {\cal G}_2^\#$ and $\omega_p>0$. Then
\begin{align*}
\theta_g'(\omega_p)  > \varrho(\omega_p)
~~\Rightarrow~~ 
(\ref{exactRIR})
~~\Rightarrow~~ 
\theta_g'(\omega_p)  \ge \varrho(\omega_p),
\end{align*}
where $\varrho(\omega):=\left|\sin(\theta_g(\omega))/\omega\right|$.
\item[(III)] For any $g \in {\cal G}_1^\#$, we have $\rho_*(g) > 1/\|g\|_{L_\infty}$. 
\end{itemize}
\end{theorem}
For the proofs of these results, readers are referred to Section 4 of~\citep{HKSZIH:arXiv2022}. 
Also note that, the necessary conditions in statements (I) and (II) hold in fact for systems in
$\mathcal{G}_n^0$ and $\mathcal{G}_n^{\#}$, respectively, for any $n$. 

\begin{table*}[h]
\caption{Summary of the numbers of peak-gains, satisfaction of the PCR conditions, whether exact RIR occurs, etc. among different cases.}
\begin{center}
  \begin{tabular}{|l||c|c|c|c|c|c|}
    \hline
    \multirow{2}{*}{} & \multicolumn{6}{|c|}{$m$} \\
    \cline{2-7} 
     & $1-4$ & $5$ & $6-7$ & $8-13$ & $14-16$ & $17-20$      \\
  \hline\hline
   {\# of unstable poles} & $2$  & $2$ & $2$  & $4$  & $4$  & $4$ \\
  \hline 
   \# of peak-gains & $1$ & $2$ & $2$ & $2$ & $3$  & $3$\\
  \hline
   {\# of unstable peak-gains} & $1$ & $1$ & $1$ & $2$ & $2$  & $2$\\
  \hline
   {\# of stable peak-gains} & $0$ & $1$ & $1$ & $0$ & $1$  & $1$\\
  \hline
   global peak-gain is (s./us.)? & us & us & s & us & us  & s\\
  \hline
   PCR holds at global peak? & y & y & n & y & y  & n\\
  \hline   
   PCR holds at a local peak? & n/a & n & y & y & y  & y\\
  \hline
   RIR $=1/\|g_m\|_{L_\infty}$ ? & y & y & n & inc & inc & n\\
  \hline
   RIR $> 1/\|g_m\|_{L_\infty}$ ? & n & n & y & inc & inc & y\\
  \hline
\end{tabular} \\
{Abbreviation: 's.' -- stable; 'us.' -- unstable; 'y' -- yes; 'n' -- no; 'n/a' -- not applicable; 'inc' -- inconclusive}
\end{center}
\label{tab:summary}
\medskip
\end{table*}

\begin{figure*}[h]
\centering
\includegraphics[width=15cm]{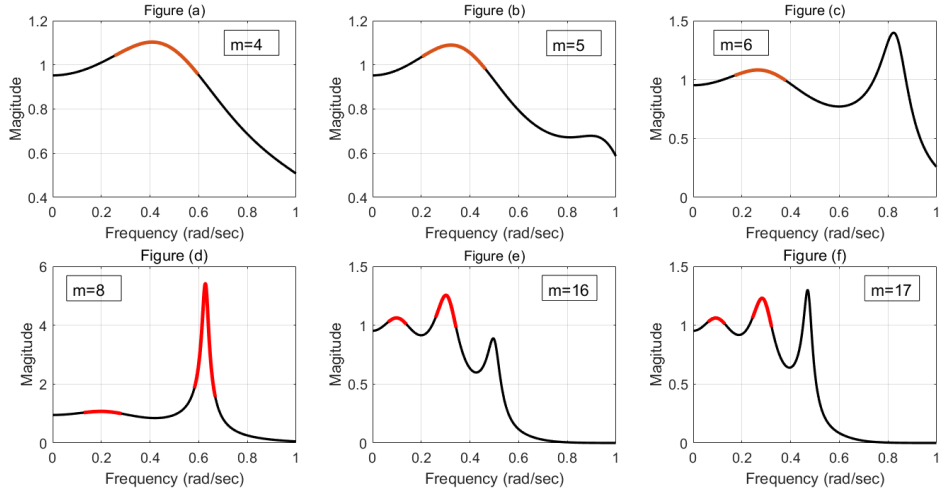}
\caption{Magnitude profile of $g_m$ for $m=4,5,6,8,16,17$. For $m=4$ to $6$, $g_m$ has one pair of unstable poles, while it has two pairs
for the other three cases. The red color indicates the frequency ranges where the PCR condition holds. A gain-peak where the 
PCR condition does not hold appears to be caused by a pair of stable poles.}
\label{fig:2.3.1} 
\medskip
\end{figure*}

\subsection{An Illustrative Example}

In this subsection we illustrate, by a numerical example, how the PCR condition effectively
works for the robust instability analysis. Consider a class of positive feedback systems of which the loop 
transfer functions are represented by 
 $h(s) = -k/(s+1)^{2m+1}, \; \; m=1, 2, \dots$,
where we assume that the loop-gain $k>0$ is large enough so that the closed-loop system is exponentially unstable. 
Our interest here is to assess robust instability against a ball type multiplicative stable perturbation;
in other words, the perturbed system $\tilde{h}$ has the form
$\tilde{h}(s) = (1 + \delta(s)) h(s)$,  $\delta(s)  \in \RHinf$. 
Such a setting may arise when one considers a cyclic network with $2m+1$ identical agents with a multiplicative uncertainty
present for the loop. The corresponding characteristic equation of the closed-loop system is given by 
$1 - g_m(s)\delta(s) = 0$, where 
$g_m(s) := h(s)/(1-h(s)) = -k/\left((s+1)^{2m+1}+k\right)$.   
For $k=20$, we observe that $g_m\in\mathcal{G}_2^{\#}$ for $1\le m\le 7$, 
and $g_m\in\mathcal{G}_4^{\#}$ when $8\le m \le 20$. The unstable poles of $g_m$ increases further when $m$ becomes bigger. 
Table~\ref{tab:summary} summarizes the findings for $m=1$ to $20$. 

For $1\le m \le 4$, $g_m$ has one peak gain, while $g_5$ has two peak gains. In all these cases, 
the PCR condition stated in Theorem~\ref{thm:marginalstab} holds at the global peak 
frequencies. See Fig.~\ref{fig:2.3.1}(a) and~\ref{fig:2.3.1}(b) for an illustration of the magnitude profiles
of $g_4$ and $g_5$. 
For $g_5$, applying Proposition~\ref{thm:PCR_Sol} we obtain the first-order all-pass function
of the form $\delta_{\rm gl,5}(s)=\frac{1}{1.0896}\left(\frac{s-24.426}{s+24.426}\right)$, which marginally stabilizes $g_5$ and 
the closed-loop system has a pair of poles at $\pm j\omega_p = \pm j(0.322)$. In this case, we conclude 
that $g_5$ has the exact RIR equal to $1/|g_5(j(0.322))| = 1/1.0896$. 

For $m=6, \ 7$, the PCR condition fails at the global peak frequencies for $g_m$. However for each case, there is
a local peak frequency where the PCR holds. See Fig.~\ref{fig:2.3.1}(c) for an illustration of the 
magnitude profile of $g_6$.
Further examination reveals that the global peak-gain is
due to a pair of dominating stable poles, while the local peak-gain is the result of a pair of unstable poles
which is further away from the imaginary axis compared to the dominating stable poles. Take $g_6$ for example. 
Applying Proposition~\ref{thm:PCR_Sol} at the global and local peak frequencies, we obtain first-order all-pass functions
$\delta_{\rm gl,6}(s)=\frac{1}{1.3976}\left(\frac{-s+1.2522}{s+1.2522}\right)$ and $\delta_{\rm lc,6}(s)=\frac{1}{1.0811}\left(\frac{s-18.02}{s+18.02}\right)$, respectively. The closed-loop system with $\delta_{\rm gl,6}$ is exponentially unstable, which has two unstable poles 
and two imaginary-axis poles. It appears that $\delta_{\rm gl,6}$ pushes the dominating stable poles to the imaginary axis while leaving 
the unstable poles in the ORHP. On the other hand, the closed-loop system with $\delta_{\rm lc,6}$ is marginally stable 
with a pair of poles at $\pm j\omega_p = \pm j(0.276)$. In this case, $g_6$ does not have exact RIR, and 
$\rho_*(g_6)\in ( 1/1.3976, 1/1.0811]$. Note that $\rho_*(g_6)$ is strictly larger than $1/\|g_6\|_{L_\infty}=1/1.3976$, 
as the necessary condition stated in statement (II) of Theorem~\ref{thm:exactRIR} is violated. 

For $8\le m\le 13$, $g_m$ has two peak-gains and both are caused by unstable poles. The PCR condition holds
at both peak frequencies. 
For $14\le m \le 16$, a third peak is formed, which is caused by a pair of stable poles. 
The PCR of $g_m$ is negative at this peak (let's call it a ``stable peak''). 
For $17\le m \le 20$, the stable 
peak overtakes the other two peaks and becomes the global peak. 
See Fig.~\ref{fig:2.3.1}(d) to~\ref{fig:2.3.1}(f) for an illustration of the magnitude profiles of 
$g_8$, $g_{16}$ and $g_{17}$.
Now consider $g_8$. The first-order all-pass functions
obtained by the global and local peak frequencies are $\delta_{\rm gl,8}(s)=\frac{1}{5.4116}\left(\frac{s-2.749}{s+2.749}\right)$
and $\delta_{\rm lc,8}(s)=\frac{1}{1.073}\left(\frac{s-29.498}{s+29.498}\right)$, respectively. The closed-loop system with $\delta_{\rm gl,8}$
is exponentially unstable; apparently $\delta_{\rm gl,8}$ pushes a pair of unstable poles to the imaginary axis while leaving the
other pair in the ORHP. Similar to $g_6$, $\delta_{\rm lc,8}$ is able to marginally stabilize $g_8$, and therefore we have
$\rho_*(g_8)\in [ 1/5.4116, 1/1.073]$. Note that we cannot yet exclude the possibility that $\rho_*(g_8)=1/5.4116$ since no 
necessary condition is violated. For $g_9$ to $g_{16}$, we have similar results, where the inverse of the $L_{\infty}$-gain of 
$g_m$ gives a lower bound and the second peak-gain of $g_m$ gives an upper bound. For $g_{17}$ to $g_{20}$, the situation is slightly
different. For those systems, their PCRs at the global peak frequencies violate the necessary condition for having
exact RIR's. Therefore, we know that $\rho_*(g_m)$ is strictly larger than $1/\|g_m\|_{L_\infty}$, for $m=17,\cdots,20$. For each
of these system, an upper bound for $\rho_*$ is obtained using their respective third peak-gains. 

\section{Practical Applications}
\label{sec:PracticalApp}

In this section, we apply our main results to analyze (in)stability properties of system models that are derived from
real-world applications. In Section~\ref{sec:meglev_sys} we consider linearized models for magnetic levitation systems. These models
belong to the class ${\cal G}_1^0$. In Section~\ref{sec:repressilator} we consider linearized models for a certain gene regulatory network
called ``repressilator''. These models belong to the class ${\cal G}_2^{\#}$. The goal is to illustrate that our results are
applicable to real applications to provide useful information. 

\subsection{Strong Stabilization for Magnetic Levitation Systems}
\label{sec:meglev_sys}
A typical linearized model for the magnetic levitation system~\citep{Namerikawa2001} at an equilibrium is a third-order system 
of the following form 
\begin{align*}
g(s) = k/\left((-s^2+p^2)(\tau s+1)\right),
\end{align*} 
where the pair of poles at $\pm p$ is due to the mechanical aspect of the system 
while the stable pole at $-\tau^{-1}$ comes from the electrical part. Typically, we have $\tau^{-1}\gg p$, and 
if this is the case one may assume that the factor $(\tau s+1)$ can be neglected from the dynamical model 
for control design purpose.
Here we will show that, however, there is a fundamental difference between the second- and the third-order models in terms of  
minimum-norm strong stabilization. First, consider the reduced second-order model $g_r(s)=k/(-s^2+p^2)$. One can readily
verify that $g_r\in\mathcal{G}_0^1$ with $\theta_{g_r}'(0)=0$. Despite that $g_r$ does not satisfy the sufficient
PCR condition stated in Theorem~\ref{thm:exactRIR}, we have
\begin{align}
\label{inf_gr}
\inf_{c\in\IS(g_r)}\|c\|_{H_\infty} = p^2/k = 1/|g_r(0)| = 1/\|g_r\|_{L_\infty},
\end{align}
The infimum in~\eqref{inf_gr} is obtained by the stabilizing controller
$c_{\epsilon}(s)=p^2/k+\epsilon(s+z)/(s+d)$ with $0<z<d$ and arbitrarily small positive $\epsilon$.

On the other hand, for the third-order model $g$, we have 
\begin{align}
\label{lb_inf_g}
\inf_{c\in\IS(g)}\|c\|_{H_\infty} > p^2/k = 1/|g(0)| = 1/\|g\|_{L_\infty}. 
\end{align}
The strict inequality in~\eqref{lb_inf_g} is due to the fact that $g\in\mathcal{G}_1^0$ and $\theta_g'(0)=-\tau<0$,
and thus $g$ violate the necessary condition for having the  RIR by Theorem~\ref{thm:exactRIR}.

For obtaining an upper bound of the infimum, let us introduce a phase-lead compensator to raise the PCR of $g$ at 
the zero frequency. 
Consider $f(s)=\left((\tau_c+\tau)s+1\right)/(\tau_c s + 1)$ and $g_c(s)=g(s)f(s)$. The compensated plant $g_c$ satisfies
$\theta'_{g_c}(0)=0$ for any $\tau_c>0$. This can be readily verified by checking the imaginary part of 
$\frac{d}{d\omega}\log(g_c(j\omega))$ at the zero frequency. Furthermore, we have 
$g_c\in\mathcal{G}_1^0$ if and only if $\tau_c\le 1/(p^2\tau)$. This can be shown by computing the real part of 
$\frac{d}{d\omega}\log(g_c(j\omega))$, which reveals that
\begin{itemize}
\item $\mathrm{Real}\left(\frac{d}{d\omega}\log(g_c(j\omega))|_{\omega=0}\right)=0$;
\item when $\tau_c\le 1/(p^2\tau)$, $\frac{d}{d\omega}\log |g_c(j\omega)| <0$ for any $\omega>0$; 
\item when $\tau_c > 1/(p^2\tau)$, $\frac{d}{d\omega}\log |g_c(j\omega)| >0$ for $\omega\to 0^+$.
\end{itemize}
and hence the claim. Setting $\tau_c = 1/(p^2\tau)$, we have the following result.
\begin{prop}
The compensated plant $g_c$ satisfies
\begin{align}
\label{inf_gc}
\inf_{c\in\IS(g_c)}\|c\|_{H_\infty} = 1/|g_c(0)| = 1/\|g_c\|_{L_\infty} = p^2/k,
\end{align}
which in turn implies 
\begin{align}
\label{lbub_inf_g}
1 < \inf_{c\in\IS(g)}\frac{\|c\|_{H_\infty}}{p^2/k} \le (1+p^2\tau^2). 
\end{align}
\end{prop}
\begin{proof} The infimum in~\eqref{inf_gc} is obtained by the stabilizing controller
$c_{\epsilon}(s)=p^2/k+\epsilon(s+\epsilon^2)/\left(s+q/(\tau_c+\tau)\right)$.
One can verify that the characteristic equation of the closed-loop system $[g_c, c_\epsilon]$ has the 
form $s^5 + [(q+1)d] s^4 + [q d^2] s^3 + [k\epsilon d] s^2 + [k\epsilon(\hat{d}+\epsilon^2 d)] s + [k\epsilon^3 \hat{d}]$, where 
$d:=(\tau+\tau_c)/(\tau\tau_c) = \tau^{-1}+p^2\tau$, and $\hat{d}:=1/(\tau\tau_c)=p^2$. The goal here is to select parameters
$\epsilon>0$ and $q>0$ such that the roots of the polynomial are all in the open left-half plane. Applying the Routh-Hurwitz 
stability criterion, one concludes that it is so when $\epsilon$ is sufficiently small and, corresponding to an $\epsilon$, $q$
is chosen sufficiently large. The infimum in~\eqref{inf_gc} is obtained by taking $\epsilon\to 0$. Furthermore, the analysis
implies that $fc_{\epsilon}$ is a stabilizing controller for $g$. Since $\|fc_{\epsilon}\|_{H_\infty}\to p^2(1+p^2\tau^2)/k$ 
as $\epsilon\to 0$, it implies $p^2(1+p^2\tau^2)/k$ is an upper bound for $\inf_{c\in\IS(g)}\|c\|_{H_\infty}$. With~\eqref{lb_inf_g},
we hence conclude the inequalities in~\eqref{lbub_inf_g}.
\end{proof}

\begin{remark}
Since $\tau^{-1}\gg p$, we have $1+p^2\tau^2 \approx 1$. That is, the upper bound on the norm of the minimum-norm strong stabilizing
controller is very close to the lower bound $p^2/k$. 
\end{remark}

\subsection{Robust Instability Analysis for Repressilator}
\label{sec:repressilator}

Consider a biological network oscillator called the repressilator
with three dynamical units in a cyclic loop~\citep{Elowitz2000}.
{\SH
Its linearized model is the positive feedback system} with a loop transfer function
$h(s)$ represented by
$$h(s) = -k/\left((s+\alpha_1)(s+\alpha_2)(s+\alpha_3)\right),$$ 
where {\yh $k>0$}
For more details about the repressilator model, see~\citep{HIH:Automatica2021}.
Here we are interested in assessing robust instability against a ball type multiplicative stable perturbation
when the nominal dynamics are further complicated by time-delay.
We use the fifth-order Pad\'{e} approximation for the time-delay
in order to keep the model rational. Let $D^{\tau}(s)$ denote the Pad\'{e} approximation of the
time-delay transfer function $\mathrm{e}^{-\tau s}$. The corresponding
characteristic equation is $1-\delta(s) {\yh g(s)} = 0$, where
\begin{align*}
{\yh g(s)} = {\yh h(s)}D^{\tau}(s)/\left(1- {\yh h(s)}D^{\tau}(s)\right)
\end{align*}
{\SH
and the nominal system with the characteristic equation  $1={\yh h(s)}D^{\tau}(s)$ is exponentially unstable.
}

{\yh We consider the case where the parameters are $\alpha_1=0.4621$, $\alpha_2=0.5545$, $\alpha_3=0.3697$, %
and $k=2.216$. We assume that the gain $k$ does not depend on the equilibrium state of the original nonlinear system. In other words, the DC-gain of the perturbation is assumed to be zero. 
For this case, the exact RIR was calculated when $\tau=0$ in \cite{HIH:Automatica2021}. 
Hence,} in what follows, we examine the effect of the time-delay on the exact RIR. 

Numerical computations show that ${\yh g} \in{\cal G}_2^{\#}$ for $\tau\in[0,4.771]$. 
The PCR condition holds at the peak-gain frequency of ${\yh g}$ up to $\tau=3.481$, and ceases
to hold when $\tau=3.482$. Thus, ${\yh g}$ has exact RIR for $\tau\in[0,3.481]$. Furthermore, one can verify that 
when $\tau$ is large enough, a pair of stable poles of ${\yh g}$ creates a gain-peak. When $\tau=3.482$, this ``stable peak'' 
becomes dominant and the PCR condition ceases to hold at the global peak frequency. However, the condition holds
at the local (second) peak frequency. %
More specifically, when $\tau=3.482$, $\|{\yh g}\|_{L_\infty} = |g(j1.5009)| =1.10273$, while a 
local peak occurs at {\yh $\omega=0.396$} with $|g(j 0.396)|=1.10268$. The first-order all-pass function 
$\frac{1}{1.10268}\left(\frac{s-18.8246}{s+18.8246}\right)$, obtained by applying Proposition~\ref{thm:PCR_Sol} to 
the local peak frequency, marginally stabilizes ${\yh g}$. Thus, we conclude that $ 1/1.10273 < \rho_*({\yh g}) \le 1/1.10268$ when
$\tau=3.482$.

\begin{figure}[h]
\epsfig{file=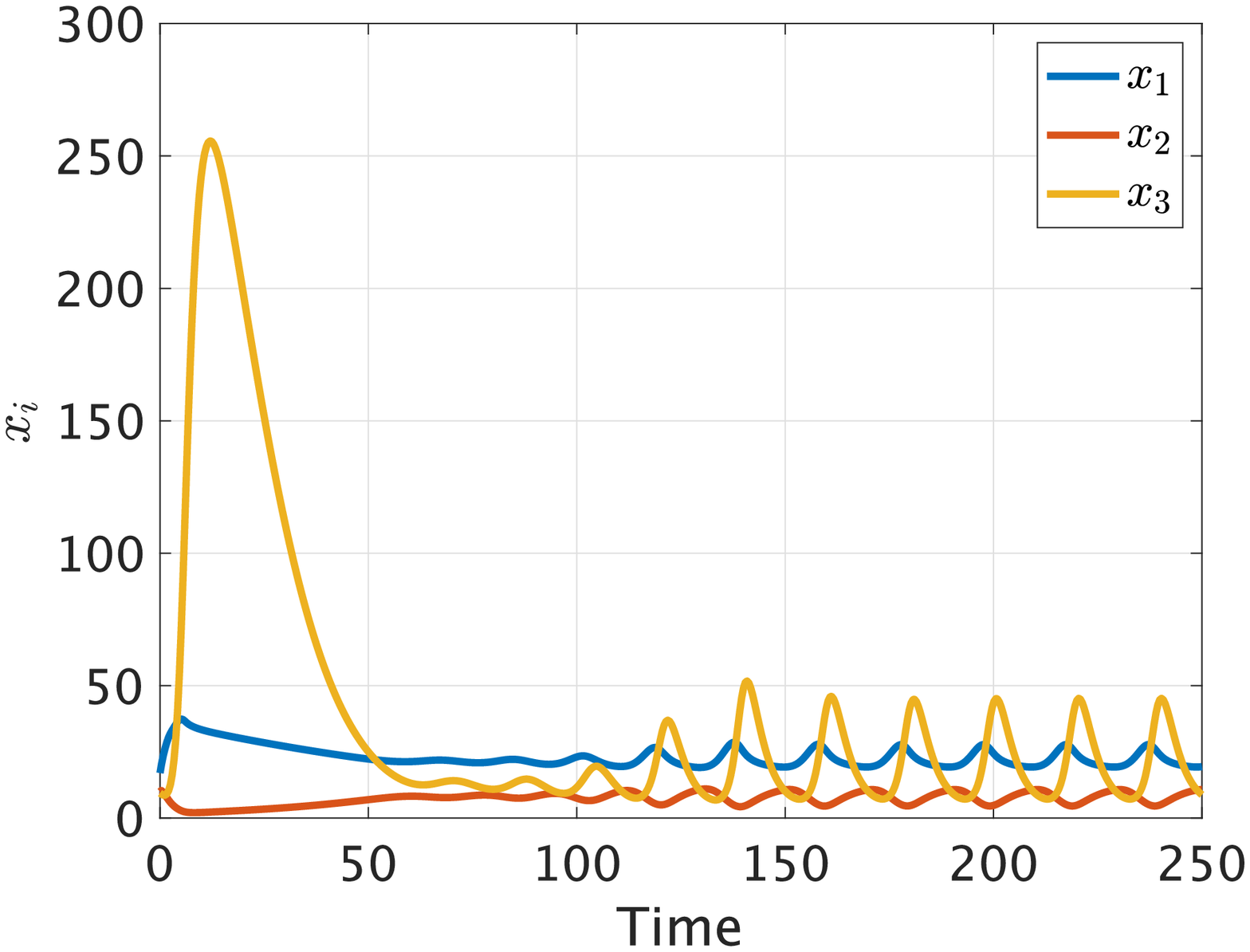, width=42.5mm}
\epsfig{file=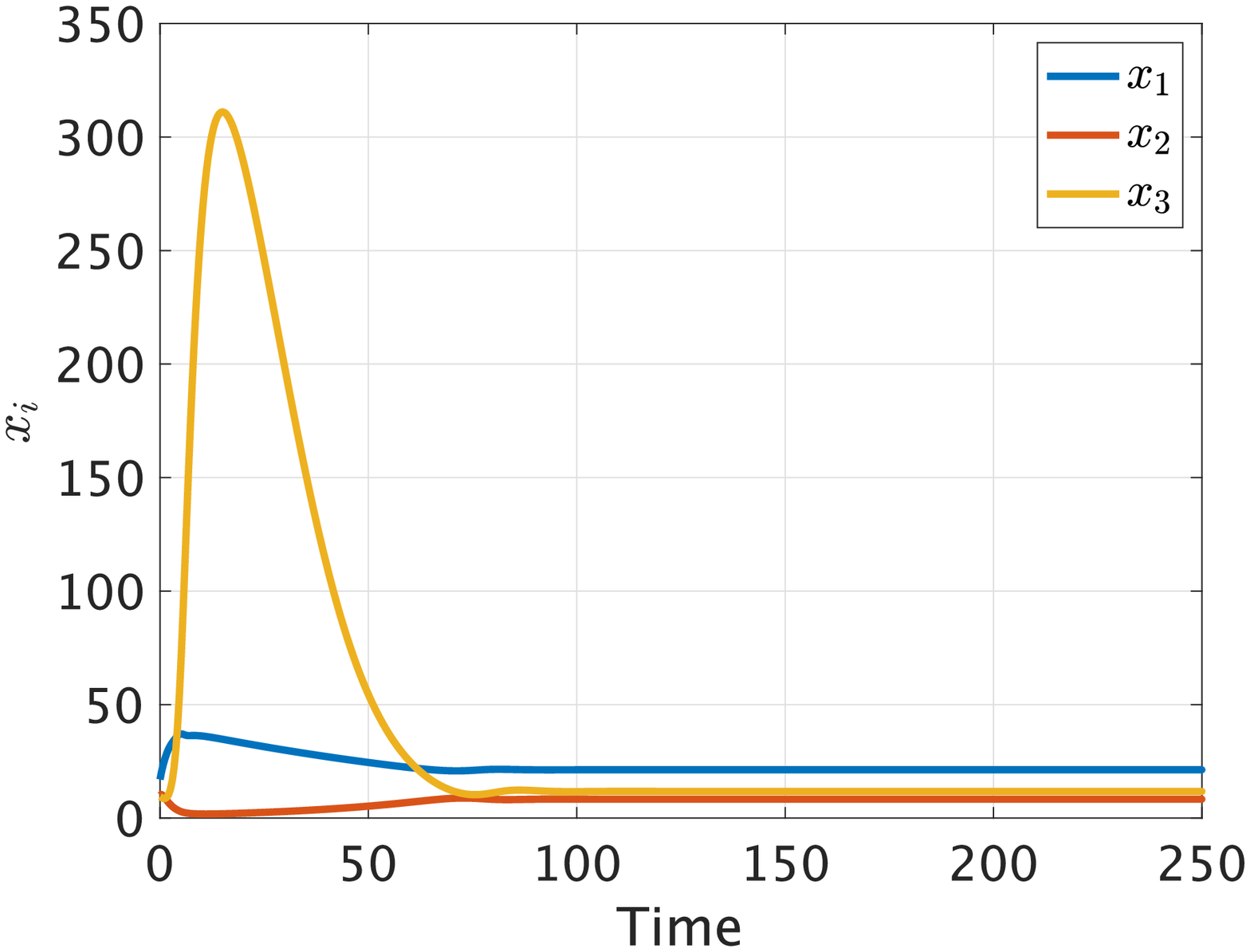, width=42.5mm}
\caption{Time-course simulations of the closed-loop systems. Left: ${\yh g}$ and $\delta(s)$ with $\epsilon={\yh -0.05}$. Right: ${\yh g}$ and $\delta(s)$ with $\epsilon = {\yh 0.05}$.}
\label{fig:3.2.2}
\end{figure}

For $\tau=3.4$, a marginally stabilizing perturbation with norm equal to $1/\|{\yh g}\|_{L_\infty}$ is
$\frac{1}{1.1044}\left(\frac{s-18.4747}{s+18.4747}\right)$. 
This perturbation is further multiplied by a high-pass filter to make the DC-gain of $\delta(s)$ equal to zero.
Specifically, $\delta(s)$ is defined by 
\begin{align*}
\delta(s) = \frac{{\yh s}}{s + {\yh 0.01}}  \cdot  (1 + \epsilon)
\frac{1}{1.1044}\left(\frac{s-18.4747}{s+18.4747}\right), 
\end{align*}
where $\epsilon$ is a real number.
The closed-loop systems of ${\yh g}$ is marginally stabilized with $\epsilon = 0$. 
The nonlinear repressilator models with $\epsilon = {\yh -0.05}$ and $\epsilon = {\yh 0.05}$
were simulated, and the results are shown in Fig.~\ref{fig:3.2.2} (left and right figures, 
respectively). Clearly, $\delta(s)$ with $\epsilon = {\yh -0.05}$ is not able to stabilize ${\yh g}$ and the closed-loop system exhibits oscillatory 
behavior. On the other hand, $\delta(s)$ with $\epsilon = {\yh 0.05}$ stabilizes ${\yh g}$ and the oscillatory behavior ceases to exist. 

\begin{remark}
In E. coli cells, the delay factor mainly represents the protein maturation time, which is usually 6 to 60 minutes. 
For the repressilator model presented in this section, the unit of time is ``hour''; therefore, the delay time 
$\tau$ of the range $[0.1,1]$ corresponds to realistic scenarios. Our analysis shows that 
the $L_{\infty}$-norm of ${\yh g}$ gives the exact RIR for $\tau\in[0,3.481]$ , which indicates
that it is a useful metric for determining the instability (i.e., oscillation) of practical repressilators.
\end{remark}

\section{Concluding Remarks}
\label{sec:Concl} 

We recalled the phase change rate maximization problem and solution from~\cite{HKSZIH:arXiv2022} and illustrated the latter's 
utility in the robust instability analysis of a cyclic network of homogenous multi-agent systems subject to an identical 
multiplicative stable perturbation on each agent. We also applied the result to two practical applications --- magnetic 
levitation systems and repressilators with time-delay. An interesting future research direction involves examining the robust 
instability of a cyclic network subject to heterogeneous multiplicative perturbations on the agents.

\end{document}